# 3D Growth and Remodeling Theory Supports the Hypothesis of Staphyloma Formation from Local Scleral Weakening under Normal Intraocular Pressure

Fabian A. Braeu[1,2,3], Stéphane Avril[4], and Michaël J.A. Girard[1,5,6,7,8,9]

1. Ophthalmic Engineering & Innovation Laboratory, Singapore Eye Research Institute, Singapore National Eye Centre, Singapore
2. Singapore-MIT Alliance for Research and Technology, Singapore
3. Yong Loo Lin School of Medicine, National University of Singapore, Singapore
4. Mines Saint-Etienne, Université Jean Monnet, Saint-Etienne, France
5. Duke-NUS Graduate Medical School, Singapore
6. Department of Biomedical Engineering, College of Design and Engineering, National University of Singapore
7. Department of Ophthalmology, Emory University School of Medicine, Atlanta, GA, USA
8. Department of Biomedical Engineering, Georgia Institute of Technology/Emory University, Atlanta, GA, USA
9. Emory Empathetic AI for Health Institute, Emory University, Atlanta, GA, USA

**Corresponding Author:**     Michaël J.A. Girard
Ophthalmic Engineering & Innovation Laboratory (OEIL)
Singapore Eye Research Institute (SERI)
The Academia, 20 College Road
Discovery Tower Level 6,
Singapore 169856
mgirard@ophthalmic.engineering
https://www.ophthalmic.engineering

# Introduction

Myopia is a refractive error characterized by blurring of objects viewed at a distance. It is commonly caused by an excessive axial growth of the eye, causing abnormal projection of remote images in front of the retina [1]. Myopia is one of the most prevalent eyes diseases worldwide [2], and its prevalence can rise up to alarming magnitudes (>80%) in East and Southeast Asian countries [3]. Adults with myopia and excessive axial length are at a high risk of developing pathologic myopia, a severe form of myopia that is one of the leading causes of irreversible vision loss [4]. Pathological myopia is defined as high myopia in the presence of additional structural changes including, but not limited to, posterior staphylomas [5].

A posterior staphyloma has been defined as an ectatic outpouching of the posterior sclera, with a characteristic change in scleral curvature at the transition zone (staphyloma ridge) and additional scleral thinning [6]. Posterior staphylomas have been grouped in 10 types, depending on their size, shape, and location, the 5 major ones being shown in **Figure 1** [7]. The presence of posterior staphylomas often leads to the development of several pathologic features such as myopic traction maculopathy (e.g. foveoschisis, macula hole, retinal detachment), choroidal neovascularization, and chorioretinal atrophy, making early detection and treatment of this condition crucial for protecting the patient's vision [6].

Despite ongoing research, the underlying causes of posterior staphyloma formation and progression remain largely unknown. Various theories have been proposed, suggesting that the sclera, choroid, or Bruch's membrane may play a role in the development of this condition [6]. However, conclusive evidence to support these hypotheses is currently lacking. As a result, our understanding of the pathophysiology of posterior staphylomas is limited, hindering our prognostic abilities for their formation, and the development of effective and targeted treatment options for individuals affected by this condition.

In recent years, mathematical models have been developed to predict the growth (change in volume) and remodeling (change in microstructure) of soft tissues in response to mechanical stimuli through a process known as mechanotransduction [8]. These Growth & Remodeling (G&R) models have been effectively used to gain a deeper understanding of the formation of aortic aneurysms [9, 10], which share similarities with staphylomas. By utilizing these models on patient-specific 3D geometries of the eye (as obtained with optical coherence tomography [OCT] or magnetic resonance imaging [MRI]), we believe we could provide a comprehensive understanding of the pathology of staphylomas but also aid in the prediction of their formation and progression. To date, it is important to emphasize that, there are currently no reliable methods or tools available for predicting the formation or progression of staphylomas. Novel tools are urgently needed.

The aim of this study was to use G&R theory to explore a potential mechanism behind the development of posterior staphylomas. Specifically, we wanted to assess whether G&R theory could explain staphyloma formation from a local scleral defect – as could occur from age-related elastin degradation, asymmetric myopia progression, or other contributing factors. This hypothesis was tested under conditions of normal intraocular pressure (IOP) since evidence suggests that the onset of myopia and staphyloma typically occurs in the absence of elevated IOP [11].

# Methods

For this work, we created a finite element (FE) model of the posterior pole of the eye and leveraged a well-established mathematical model of G&R, namely the homogenized constrained mixture model (HCMM) [10, 12], to simulate structural adaptations over time resulting from a localized scleral weakening.

**Finite element model of the eye**

**Major assumptions.** The herein developed FE model to study posterior staphyloma formation relies on several assumptions:

1) The sclera and lamina cribrosa (LC) serve as the main load-bearing tissues of the eye. While we acknowledge that the choroid and Bruch's membrane may contribute to the structural strength of the eye, for the purpose of this study, we assumed that this contribution is negligible. Consequently, we only modeled the sclera and LC.

2) Staphyloma formation is mechanically driven, unlike myopia, where there is an additional optical component [11].

3) Neural tissues were excluded from the model, as they are not believed to play a role in staphyloma formation [11].

4) IOP remains within normal range, consistent with clinical observations in myopia and staphyloma [11].

5) Local scleral weakening serves as trigger for staphyloma formation, which may occur in adulthood through elastin degradation [13] or myopic growth [14].

**Eye geometry.** To investigate the development of posterior staphyloma driven by G&R processes, we focused our investigation on collagen-rich connective tissues like the LC, the peripapillary sclera (PPS), and the peripheral sclera (PS). For this purpose, we modelled the posterior pole of the eye (**Figure 2a**) as a semi-spherical shell divided into three distinct regions along the meridional direction at 4 and 30 degrees representing the LC, the PPS, and the PS (**Figure 2b**). The spherical shell had an inner radius of 12 mm and a uniform thickness of 0.5 mm. For computational efficiency, we limited our simulation to a 3-degree sector of the semi-spherical shell (**Figure 2c**). Through the utilization of rotational symmetry boundary conditions, this approach effectively simulated the entire semi-spherical shell. Based on a convergence study to ensure negligible numerical errors, we discretized the geometry with 900 eight-node hexahedral (Abaqus: C3D8R) and 30 six-node wedge elements (Abaqus: C3D6).

**Boundary conditions.** The equator of the eye model was fixed along the normal direction to the surface (**Figure 2d**, y direction), allowing for free expansion of the equator in the transverse direction (**Figure 2d**, x and z direction). Furthermore, rotational symmetry boundary conditions were imposed on the two surfaces in circumferential direction. A uniform pressure load of 15 mmHg (i.e. IOP) was applied to the inner surface of the semi-spherical shell.

**Mechanical constitutive equations.** We modelled each tissue, i.e. the LC, PPS, and the PS, as a constrained mixture of multiple constituents i, such as collagen and elastin, that deform together with no relative motion between them. Therefore, in the HCMM, the total deformation gradient was multiplicatively split into

$$\mathbf{F} = \mathbf{F}_e^i \mathbf{F}_g^i \mathbf{F}_r^i \qquad (1)$$

where $\mathbf{F}_e^i$ is the elastic deformation gradient that ensures that all constituents deform together, $\mathbf{F}_g^i$ is the inelastic growth deformation gradient that describes the local change if size and shape by deposition or degradation of mass, and $\mathbf{F}_r^i$ is the inelastic remodeling deformation gradient that captures changes in the microstructure by mass turnover (i.e. the continuous deposition and degradation of tissue mass) of constituent i. The latter two were defined by mechanobiological constitutive equations that are introduced in a subsequent section.

The Cauchy stress tensor $\boldsymbol{\sigma}$ results in

$$\boldsymbol{\sigma} = \frac{2}{J}\mathbf{F}\frac{\partial \Psi}{\partial \mathbf{C}}\mathbf{F}^{\mathrm{T}} \qquad (2)$$

with the determinant of the deformation gradient $J = \det(\mathbf{F})$ and the total strain energy (per unit volume)

$$\Psi = \sum_{i=0}^{n} \rho_0^i(t) W^i(\mathbf{C}_e^i). \qquad (3)$$

$W^i$ is the strain energy per unit mass, $\rho_0^i$ is the mass per reference volume (referred to as reference mass density), $\mathbf{C}_e^i = \mathbf{F}_e^{i\,\mathrm{T}}\mathbf{F}_e^i$ is the elastic right Cauchy-Green deformation tensor of constituent i, and n is the total number of constituents. This homogenization of the stress response across all constituents is denoted as *rule-of-mixture* [15]. Note that the strain energy in equation (3) only depends on the elastic deformation. Deformation that stems from inelastic deformation (i.e. G&R) does not store energy and therefore does not directly create any stress.

Specifically, we assumed that each tissue (i.e. the LC, the PPS, and the PS) was comprised of collagen fibers organized circumferentially (i = fc) around the optic disc (**Figure 2b**), meridional collagen fibers (i = fm) extending from the posterior pole to the equator (**Figure 2b**), and a ground matrix (i = m).

To model the mechanical behavior of collagen fibers, we implemented the following strain energy per unit mass

$$W^f = \frac{c_3}{2c_4}\left\{\exp\left[c_4\left(I_{4e}^f - 1\right)^2\right] - 1\right\} \qquad (4)$$

with $f \in \{fc, fm\}$, the stress coefficient $c_3$, the uncrimping rate $c_4$, and the fourth pseudo-invariant of $\mathbf{C}_e^f$ that is defined as $I_{4e}^f = \mathbf{a}_{gr}^f \cdot \mathbf{C}_e^f \mathbf{a}_{gr}^f$. Here, $\mathbf{a}_{gr}^f$ represents the fiber direction in the intermediate configuration that is related to the reference configuration via $\mathbf{F}_g^f \mathbf{F}_r^f$. Equation (4) was adapted from [16], with the exception that we did not perform an isochoric-volumetric split of the strain energy function [17].

Motivated by the different material behaviour in shear (i.e. isochoric deformation) and bulk (i.e. volumetric deformation), we modelled the ground matrix as a decoupled (i.e. isochoric-volumetric split) isotropic Neo-Hookean material

$$W^m = \overline{W}^m(\overline{\mathbf{C}}_e^m) + K(J_e^m - 1)^2 = c_1[\text{tr}(\overline{\mathbf{C}}_e^m) - 3] + K(J_e^m - 1)^2 \qquad (5)$$

The first summand $\overline{W}^m(\overline{\mathbf{C}}_e^m)$ corresponds to the isochoric part of the strain energy function with the shear stiffness $c_1$ and the isochoric elastic Cauchy-Green deformation gradient of the ground matrix $\overline{\mathbf{C}}_e^m = (J_e^m)^{-2/3} \mathbf{C}_e^m$ with the determinant of the elastic deformation gradient $J_e^m = \det(\mathbf{F}_e^m)$. The second summand corresponds to the volumetric part of the strain energy function with bulk modulus K.

In summary, the total strain energy (per unit volume) of the constrained mixture results in

$$\Psi = \rho_0^m \overline{W}^m(\overline{\mathbf{C}}_e^m) + \rho_0^m K(J_e^m - 1)^2 + \rho_0^{fc}(t) W^{fc}(\mathbf{C}_e^{fc}) + \rho_0^{fm}(t) W^{fm}(\mathbf{C}_e^{fm}). \qquad (6)$$

The constitutive parameters were adjusted to align with experimental observations from [18, 19], and the specific values are presented in **Table 1**. For a detailed derivation of the Cauchy stress and the spatial elasticity tensor of the constrained mixture, the interested reader is referred to the **Appendix**.

**Mechanobiological constitutive equations.** To close the system of equations and enable the computation of a mechanical equilibrium state, it was necessary to define mechanobiological constitutive equations that govern the evolution of the reference mass densities $\rho_0^i$ and the G&R deformation gradients $\mathbf{F}_g^i$ and $\mathbf{F}_r^i$, in addition to the mechanical constitutive equations. In this study, all mechanobiological constitutive equations were based on the generally accepted assumption that each biological tissue has a preferred mechanical state, also known as the homeostatic state [20]. Furthermore, considering the decrease in elastin synthesis with age and the near cessation in adults, we assumed that only collagen was subject to G&R.

**Mechanobiological constitutive equations: Net mass production.** Following previous publications of aortic G&R [21, 22], we defined the net mass production rate (i.e. the difference between mass production and degradation) as

$$\dot{\rho}_0^f = \rho_0^f(t) k_\sigma \Delta G^f. \qquad (7)$$

Here, $k_\sigma$ is a growth parameter that governs the net mass production rate and $\Delta G$ represents the stress-driven growth stimulus

$$\Delta G = \frac{\sigma^f - \sigma_h}{\sigma_h} \qquad (8)$$

where $\sigma^f$ denotes the Cauchy stress of collagen fibers in their respective directions (i.e. circumferential and meridional) and $\sigma_h$ represents the collagen fiber stress at homeostasis.

**Mechanobiological constitutive equations: Growth.** In this study, we investigated two distinct types of growth wherein the net mass production is **(1)** solely linked to a change in volume (referred to as volumetric growth) or **(2)** associated with a change in mass density (referred to as mass density growth). In the following, we assumed that all constituents of the constrained mixture experience the same inelastic growth deformation gradient $\mathbf{F}_g = \mathbf{F}_g^i$, aligning with a similar assumption made in [10]. In the case of volumetric growth, we considered the special case of anisotropic volumetric growth in wall thickness direction, termed transmural growth, where the inelastic growth deformation gradient was defined as

$$\mathbf{F}_g = \frac{\rho_0(t)}{\rho_0(0)}(\mathbf{a}_0^\perp \otimes \mathbf{a}_0^\perp) + (\mathbf{I} - \mathbf{a}_0^\perp \otimes \mathbf{a}_0^\perp) \qquad (9)$$

with the total reference mass density $\rho_0(t) = \sum_{i=0}^n \rho_0^i(t)$ at current time t, the total reference mass density $\rho_0(0) = \sum_{i=0}^n \rho_0^i(0)$ at time zero, $\mathbf{a}_0^\perp$ is the transmural direction in the reference configuration, and $\otimes$ represents a tensor product. This specific choice of $\mathbf{F}_g$ resulted in a constant total mass density of the constrained mixture in the spatial configuration. In the case of mass density growth, we chose the growth deformation gradient to be equal to the identity tensor

$$\mathbf{F}_g = \mathbf{I} \qquad (10)$$

which automatically led to a change in total mass density in the spatial configuration and, consequently, a change in effective stiffness of the individual constituents.

**Mechanobiological constitutive equations: Remodeling.** In the HCCM, we assume that remodeling occurs due to continuous mass turnover. Existing mass at the current stress $\sigma^i$ degrades, and new mass is deposited at a stress equal to the homeostatic stress $\sigma_h$. This local change in mass alters the stress state of each constituent in the constrained mixture, resulting in small sliding or relative deformations between adjacent fibers and/or lamellae [23]. This phenomenon is known as remodeling. In this study, only collagen was subject to remodeling resulting in $\mathbf{F}_r^m = \mathbf{I}$.

Given this understanding of remodeling and the assumption that the collagen fiber families were modeled as quasi-1D incompressible materials, one possible definition for the remodeling deformation gradient was

$$\mathbf{F}_r^f = \lambda_r^f(\mathbf{a}_0^f \otimes \mathbf{a}_0^f) + \frac{1}{\sqrt{\lambda_r^f}}(\mathbf{I} - \mathbf{a}_0^f \otimes \mathbf{a}_0^f). \tag{11}$$

Here, $\mathbf{a}_0^f$ is the respective collagen fiber direction in the reference configuration and $\lambda_r^f$ is the inelastic remodeling stretch along $\mathbf{a}_0^f$. The inelastic remodeling stretch was determined by the following evolution equation

$$\dot{\lambda}_r^f = \left[\frac{\dot{\rho}_0^f}{\rho_0^f} - \frac{1}{T}\right][\sigma^f - \sigma_h]\left[\frac{\rho_0^f}{J}\left(\frac{\partial W^f}{\partial \lambda_e^f} + \lambda_e^f \frac{\partial^2 W^f}{\partial \lambda_e^f \partial \lambda_e^f}\right)\right]^{-1}\frac{\lambda_r^f}{\lambda_e^f} \tag{12}$$

with the elastic stretch $\lambda_e^f$ along $\mathbf{a}_0^f$ and the average turnover time T. Please note that the evolution of the inelastic remodeling stretch was driven by the difference between collagen fiber stress $\sigma^f$ and homeostatic stress $\sigma_h$, which is the same homeostatic stress as in equation (8). For a detailed derivation of equation (11) and (12), we refer the interested reader to [12].

**Predicting posterior staphyloma formation from local scleral weakening**

To investigate posterior staphyloma formation from an initially healthy state of the sclera and LC (i.e. a mechanical equilibrium configuration where no mechano-regulated G&R occur), we divided the simulation into three major steps:

(1) **Establish mechanical and mechanobiological equilibrium under normal IOP**: We prescribed the elastic stretch $\lambda_e^i$ of each collagen fiber family to equal the homeostatic stretch $\lambda_h$ or, equivalently, homeostatic stress $\sigma_h$ which are directly related through equation (2), (3), and (4). Additionally, we applied a normal IOP of 15 mmHg. Given the absence of experimental data on residual stresses in various eye tissue constituents (i.e. collagen and elastin), we did not control the elastic stretch of the ground matrix, as was done in prior work on vascular and cardiac G&R [24, 25].

(2) **Induce local scleral weakening:** We reduced the shear stiffness of the ground matrix in the PPS and LC by 85%.

(3) **Simulate G&R in the LC and PPS.** Our simulations were performed over a span of 5,000 days (approximately 13.7 years).

According to these steps, we studied three different G&R scenarios by varying the rate of mass turnover (i.e. changing the growth parameter $k_\sigma$) and the type of growth:

(1) A growth parameter of $k_\sigma = 2 \cdot 10^{-4}$ days$^{-1}$ in combination with transmural volumetric growth.

(2) An increased growth parameter of $k_\sigma = 2 \cdot 10^{-3}$ days$^{-1}$ in combination with transmural volumetric growth.

(3) A growth parameter of $k_\sigma = 2 \cdot 10^{-3}$ days$^{-1}$ in combination with mass density growth.

We implemented the HCMM as a user-defined material subroutine (UMAT) in the ABAQUS/Standard FE package [26]. The code is available on GitHub (https://github.com/fbraeu90/GR-Eye).

**G&R simulation outcome measures.** For each of the three scenarios, we recorded the maximum growth stimulus for both, circumferential and meridional fibers, in the PPS over time. Furthermore, we reported the average change in PPS thickness and the average change in proportion of circumferential and meridional collagen fibers within the PPS after 13.7 years of mechano-regulated G&R initiated by scleral weakening.

# Results

In **scenario 1** ($k_\sigma = 2 \cdot 10^{-4}$ days$^{-1}$), we observed an initial increase in collagen fiber stress within the PPS region due to a reduction of the shear stiffness of the ground matrix, resulting in an increase in growth stimulus. Subsequently, stress levels/growth stimuli decreased in an attempt to reach the homeostatic level. However, after approximately 1,500 days of G&R, we observed a tipping point where collagen fiber stresses and consequently growth stimuli began to steadily increase. This led to significant growth of the posterior pole, likely persisting without interruption, even after 13.7 years (**Figure 3**). The final shape of the simulated posterior staphyloma after 13.7 years of G&R (**Figure 4**) closely resembled that of a Type-III staphyloma, also known as peripapillary staphyloma, an example of which, captured with MRI, can be observed in **Figure 5** [27]. Additionally, we found a notable change in scleral curvature, a shift in collagen fiber distribution from a ratio of 90:10 of circumferential to meridional fibers to 75:10, and a thinning of the PPS from 0.50 mm to an average of 0.08 mm (an 84% reduction in thickness).

In contrast, **in scenario 2** and **3** where the growth parameter was increased from $k_\sigma = 2 \cdot 10^{-4}$ days$^{-1}$ to $k_\sigma = 2 \cdot 10^{-3}$ days$^{-1}$, we observed that following an initial rise in collagen fiber stress/growth stimulus in the PPS region due to the reduction of the shear stiffness of the ground matrix, the collagen fiber stress in both, the meridional and circumferential directions, gradually returned to their homeostatic level, resulting in growth stimuli close to zero (**Figure 3**). Both scenarios exhibited characteristics of a posterior staphyloma. However, the deformations were less drastic compared to **scenario 1**. In both cases, the posterior staphyloma remained stable with almost no further growth after approximately 2500 days of G&R. We found that mass density growth (**scenario 3**) resulted in a thinner PPS region (58% reduction from 0.50 mm to 0.21 mm) compared to a 30% reduction in average PPS thickness from 0.50 mm to 0.35 mm in the case of transmural volumetric growth (**scenario 2**). In both scenarios, the average collagen fiber distribution in the PPS region shifted from 90% circumferential and 10% meridional to 82% circumferential and 18% meridional.

## Discussion

In this study we used G&R theory to explore a potential mechanism behind the development of posterior staphylomas. Specifically, we found that G&R theory could explain staphyloma formation from a local scleral defect within the ground substance matrix and it did so under normal levels of IOP. While this work remains preliminary, it establishes a foundation that could ultimately guide us to identify the main mechanism behind staphyloma formation – a critical step for progress in treatment strategies.

In this study, we found that a local scleral defect could be the sole driving factor for the formation and development of posterior staphylomas. Specifically, a weakening of the ground substance matrix within the peripapillary sclera was a sufficient trigger to elicit either stable or unstable eye growth (Type-III staphylomas). Eye growth patterns highly resembled those observed with MRI [28]. Notable features of our model that were consistent with clinical observations included ectatic outpouching of the peripapillary sclera, a characteristic change in scleral curvature at the staphyloma ridge, and drastic thinning of the sclera within the staphyloma region. While our models were unable to determine the specific trigger for the weakening of the ground substance matrix, it is conceivable that this could result from either asymmetric myopic growth, or age-related elastin degradation, both of which could compromise the mechanical integrity of the cornea-scleral shell locally. This exact mechanism, however, has yet to be established.

It is interesting to note that other researchers have proposed that the underlying cause may be related to the choroid or Bruch's membrane [6]. A recent study found that 86% of eyes with staphylomas had such thin choroids in the staphyloma region that thickness could not be accurately measured using OCT imaging [29]. This extreme thinning of the choroid, which provides oxygen and nutrients to the anterior sclera, could lead to local scleral thinning and decreased resistance to IOP at the posterior pole. If the growth of staphyloma was solely driven by the sclera, it might result in choroidal thickening rather than the marked thinning that is typically observed. Another theory for the development of posterior staphyloma posits that it may be caused by excessive growth of Bruch's membrane. If this membrane were to grow excessively, it could compress the choroid against the sclera, leading to thinning of the choroid and pushing of the sclera posteriorly in specific areas of weakness. This hypothesis suggests that Bruch's membrane, rather than the sclera, may play a role in the development of staphyloma and myopia [30]. At this stage, our models cannot exclude the implication of the choroid or Bruch's membrane in the formation of posterior staphylomas.

Furthermore, we found that an increase in growth parameter $k_\sigma$ from $2 \cdot 10^{-4}$ days$^{-1}$ (in **scenario 1**) to $2 \cdot 10^{-3}$ days$^{-1}$ (in **scenario 2** and **3**) resulted in less pronounced deformations of the posterior pole of the eye, leading to a smaller posterior staphyloma after 13.7 years of G&R. Recently, a global sensitivity analysis of the HCMM confirmed the significant impact of $k_\sigma$ on G&R deformation, surpassing all other input parameters [31]. To translate this computational model to clinical practice and enable patient-specific prediction of staphyloma formation, determining a patient-specific $k_\sigma$ becomes crucial. This could potentially be achieved by following a patient over a few years and monitoring alterations in the shape of the posterior pole of the eye across multiple OCT scans. By employing the inverse finite element method, one could then derive a patient-specific $k_\sigma$ from these morphological changes, thereby enabling personalized predictions of staphyloma growth. However, the clinical feasibility of such an approach remains to be demonstrated in a longitudinal study.

In this study, we examined two distinct types of growth and found that mass density growth, where alterations in mass are linked to changes in current mass density, resulted in a thinner PPS compared to volumetric growth, where variations in mass are solely characterized by a change in volume. The prevailing understanding is that hard tissues adapt to environmental changes by increasing their density and soft tissues adapt by increasing their volume [32]. Accordingly, previous studies of G&R of hard tissues, like bones, have mainly employed mass density growth [33], while volumetric growth has been explored in soft tissues such as the LC [34], the sclera [35], arteries [36], or skin [37]. In contrast to the common understanding, Hoang et al. reported an increase in scleral mass density (i.e. mass density growth) in highly myopic guinea pig eyes relative to normal eyes [38]. However, additional experimental studies are necessary to confirm such findings in humans with staphyloma. At present, we believe that volumetric growth predominates as the primary mode of growth in staphyloma formation.

There are currently limited treatment options for posterior staphyloma. One approach is scleral reinforcement, which involves attaching a piece of donor sclera to the recipient sclera to provide additional support. This method has been shown to be effective in reducing axial length in patients with myopic macular holes, foveoschisis, and foveal retinal detachment, but its effect on staphylomas has not been studied [39]. Other procedures, such as macular buckling [40] and scleral shortening [41] for myopic traction maculopathy, have been reported to result in flattening of staphylomas, although it is unclear how long this effect lasts. Another future treatment option for staphyloma may involve scleral collagen crosslinking, a technique that uses UV light or chemicals to strengthen the collagen in the sclera [42]. However, despite the various treatments that have been proposed for staphyloma, none have yet been validated. Furthermore, the current state of our understanding of the pathophysiology of staphyloma is limited, which severely constrains our ability to develop novel and targeted treatments for this condition. Through this G&R foundation, we aim to significantly enhance our understanding of the pathophysiology of staphyloma, with the ultimate goal of improving treatment outcomes for patients suffering from this debilitating condition.

In this study, several limitations warrant further discussion. First, for simplicity, the constitutive equations and biomechanical properties chosen for the posterior sclera and for the LC are typically used to describe arteries (with 2 fiber families) but not ocular tissues. This choice resulted in an overestimation of IOP-induced strains (1$^{st}$ principal; IOP: 0 to 15 mmHg) in the peripapillary sclera (on average 6%) and in the LC (on average 7.5%), while they have been reported to be on the order of 1% in healthy human sclera [43]. Additionally, we observed numerical instabilities for certain combinations of biomechanical properties. It should be emphasized that G&R theory is complex, and our goal was to demonstrate a preliminary application in ocular tissues, which necessitated the adoption of certain simplifying assumptions. In the future, we will be able to consider more biofidelic constitutive models for ocular connective tissues that would incorporate e.g. multi-directionality of the collagen fibers [44-46] and other parameters such as collagen fiber crimp [47].

Second, elastin was incorporated within the ground substance matrix and not modeled as a separate contribution within the strain energy function. In humans, there is evidence that elastin concentration can vary dramatically from the equator to the ONH region, with a higher concentration within the peripapillary sclera and LC [48]. In the eye, elastin fibers may also contribute to tissue anisotropy with a clear elastin fiber ring surrounding the optic disc [48]. Such contributions could be considered in future models.

Third, our current model was not able to capture microstructural changes observed in posterior staphylomas including localized Bruch's membrane rupture [49], de-arrangement of collagen fibrils, a reduction in diameter of collagen fibrils, and an increase in unusual star-shaped fibrils [50]. In the future, we plan to incorporate more sophisticated constitutive models capable of capturing not only macroscopic changes, such as scleral thinning, but also alterations in the microstructure.

Fourth, we did not assign a pre-stretch to the ground matrix to regulate its elastic stretch during the establishment of mechanical and mechanobiological equilibrium. Patient-specific scleral geometries can be obtained from in-vivo imaging modalities such as OCT or MRI. However, these geometries do not represent a stress-free configuration but rather a stressed state due to loads such as IOP. Therefore, incorporating a pre-stretch of the ground matrix is necessary to achieve mechanical and mechanobiological equilibrium under normal IOP without further deformation of the geometry. We plan to incorporate such a pre-stretch of the ground matrix in future patient-specific scleral models, by adapting a recently published pre-stretch algorithm for patient-specific aortic geometries [24].

Fifth, we chose a homeostatic stretch value of 1.01 for collagen, which differs from the values commonly used in arterial G&R studies that range from 1.06 to 1.10 [24, 51, 52]. These values are typically derived from residual stretches observed in the aortic wall, as determined through ring-opening experiments [53, 54]. However, such experiments are scarce in ocular tissues and are currently only available for the porcine sclera [55]. Furthermore, a study on G&R of the lamina cribrosa suggested much lower homeostatic stretch levels compared to those commonly used in arterial G&R studies [34]. It is important to note, that the main objective of this preliminary study was to demonstrate the potential of G&R theory in explaining certain pathological growth patterns observed in the eye, such as posterior staphylomas. We hope that our study will inspire further research in this area, including experimental investigations of residual stretches in the human sclera.

Sixth, we ignored the contributions of the choroid, Bruch's membrane, and retinal tissues, but also that of other loads acting on the posterior pole such as optic nerve traction, cerebrospinal fluid pressure, orbital pressure, and ocular pulse [49]. By using the proposed G&R framework as a foundation, we hope to test a wide variety of hypotheses regarding the formation of posterior staphylomas.

Seventh, our model did not incorporate a mechanism to simulate scleral rupture. While the fast majority of posterior staphylomas stabilize over time, rare instances of excessive growth can lead to scleral rupture [56]. While the unstable and potentially uninterrupted growth observed in **scenario 1** may exhibit similarities to such a case, the lack of a rupture mechanism renders it unphysiological. Such a rupture mechanism could be included in future models.

In conclusion, our proposed framework suggests that local scleral weakening is sufficient to trigger staphyloma formation under a normal level of IOP. Our model also reproduced characteristics of Type-III staphylomas. With patient-specific scleral geometries (as could be obtained with wide-field OCT), our framework could be clinically translated to help us identify those at risks of developing posterior staphylomas, while encouraging novel treatment strategies. Finally, G&R theory suggests an important role for the sclera in the development and progression of posterior staphylomas, but it does not exclude the implication of other connective tissues, such as the choroid and Bruch's membrane.

# Acknowledgment

We acknowledge funding from **(1)** the donors of the National Glaucoma Research, a program of the BrightFocus Foundation, for support of this research (G2021010S [MJAG]); **(2)** the "Retinal Analytics through Machine learning aiding Physics (RAMP)" project that is supported by the National Research Foundation, Prime Minister's Office, Singapore under its Intra-Create Thematic Grant "Intersection Of Engineering And Health" – NRF2019-THE002-0006 awarded to the Singapore MIT Alliance for Research and Technology (SMART) Centre [MJAG]. **(3)** the "Tackling & Reducing Glaucoma Blindness with Emerging Technologies (TARGET)" project that is supported by the National Medical Research Council (NMRC), Singapore (MOH-OFLCG21jun-0003 [MJAG]).

# Figures

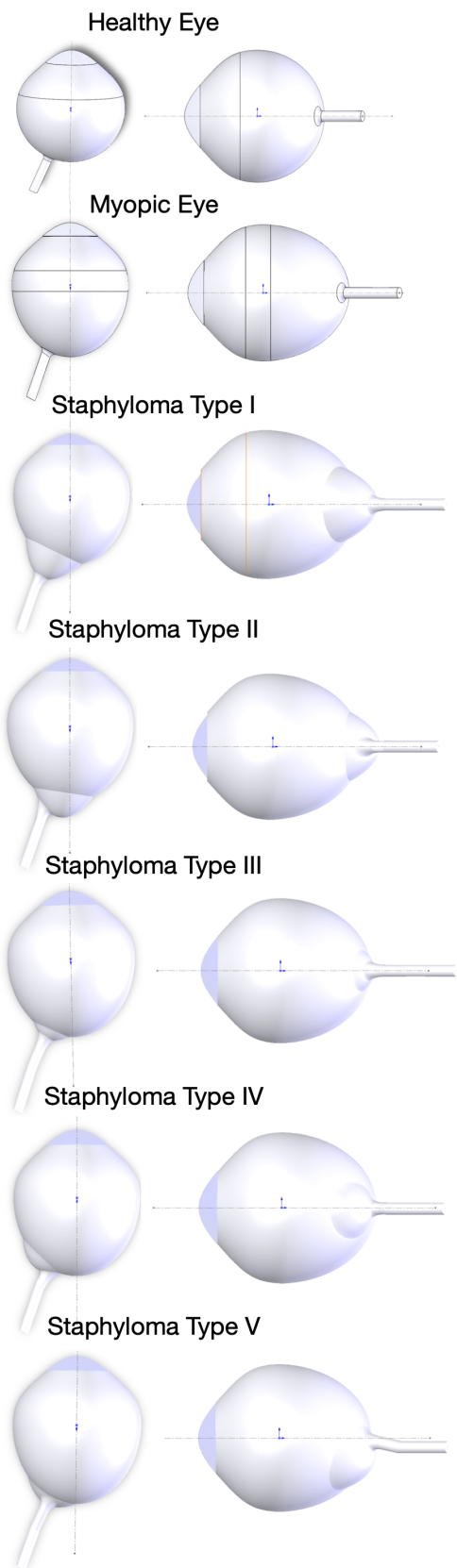

**Figure 1.** Typical morphologies of healthy eyes, highly myopic eyes, and highly myopic eyes with the presence of staphylomas (Type I-V).

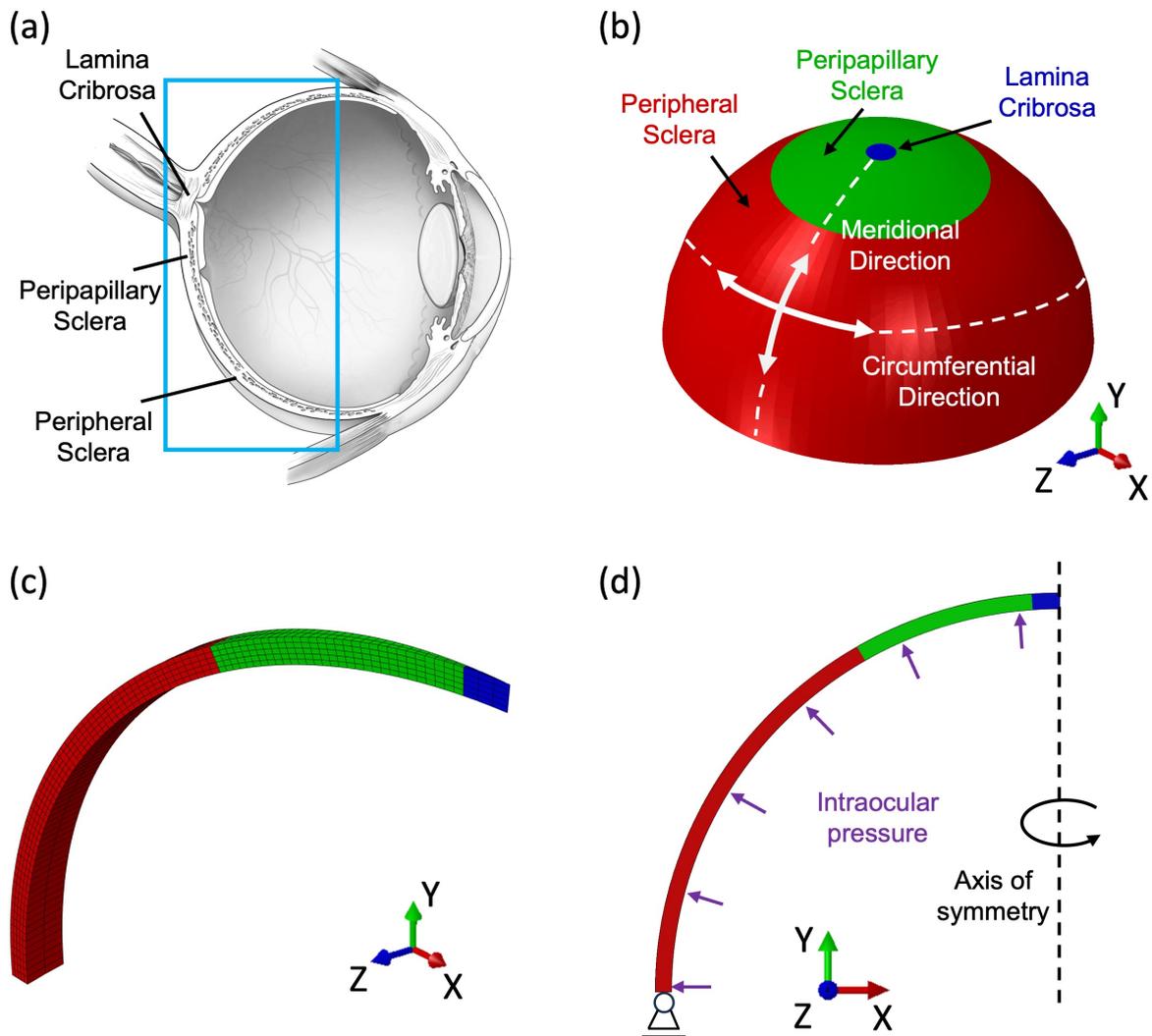

**Figure 2.** (a) Schematic of a human eye (adapted from the national eye institute: https://nei.nih.gov). The blue rectangle marks the region that was included in FE model of the posterior pole of the eye. (b) Model of the posterior pole of the eye including the lamina cribrosa, the peripapillary sclera, and the peripheral sclera. (c) 3D view of the meshed geometry used in the growth and remodeling simulations. The FE model represents a 3-degree sector of the semi-spherical shell. (d) Definition of boundary conditions.

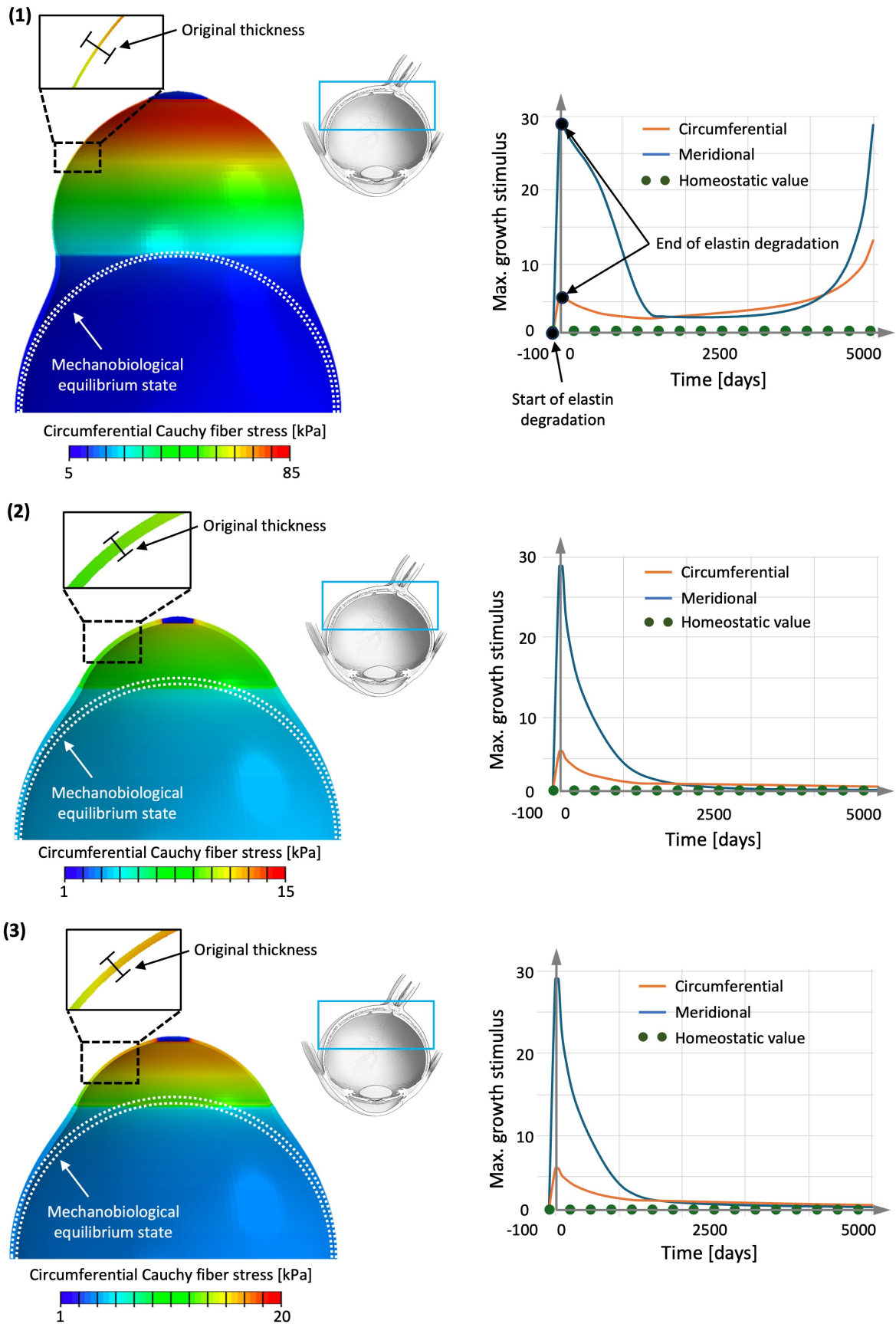

**Figure 3.** Left column: Final shape of the posterior pole of the eye after 13.7 years of G&R. **(1)**, **(2)**, and **(3)** refer to the simulated G&R scenarios. Right column: Maximum growth

stimulus (i.e. the difference between collagen fiber stress and homeostatic stress normalized w.r.t. the homeostatic stress) in the PPS over a time span of 13.7 years. Timepoint 0 represents the onset of G&R processes.

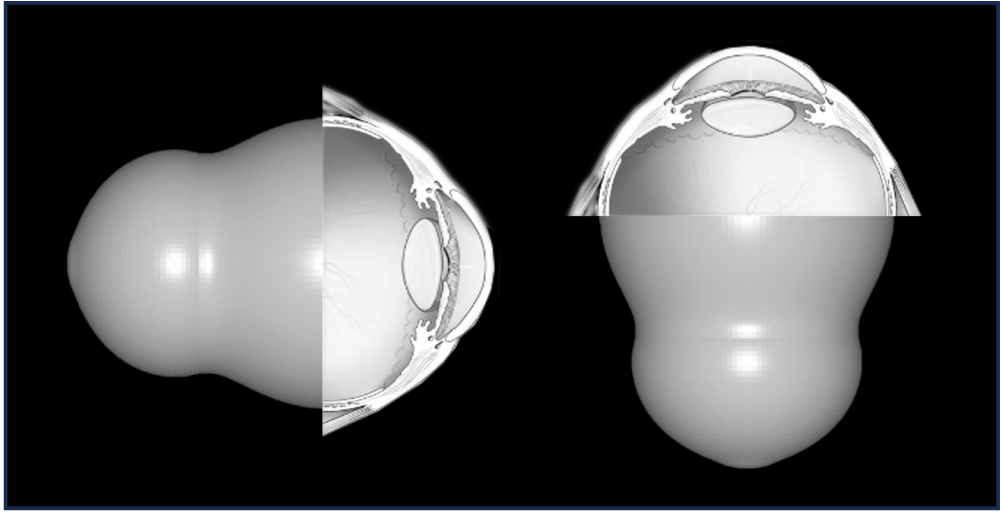

**Figure 4.** Final shape of simulated posterior staphyloma. For improved visualization, we incorporated a schematic of the anterior part of the eye (adapted from the national eye institute: https://nei.nih.gov).

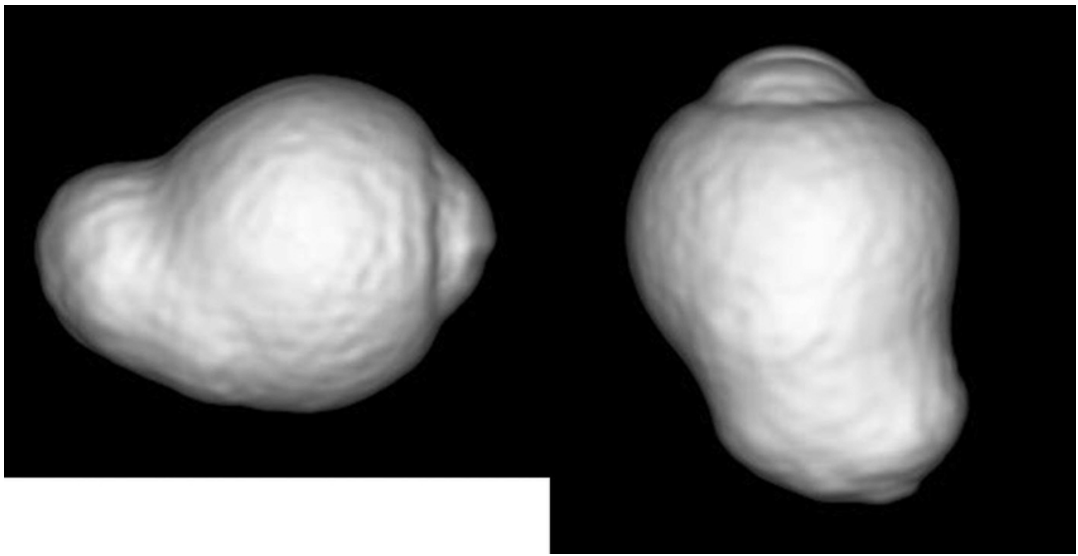

**Figure 5.** Three-dimensional MRI of an eye with a Type-III posterior staphyloma. Taken from [28] under CC BY-NC-ND 4.0 license.

# Tables

**Table 1.** Material parameters used to study posterior staphyloma formation.

| | SYMBOL | VALUE |
|---|---|---|
| **Lamina cribrosa** | | |
| Shear modulus of the ground matrix | $c_1$ | 5 J/kg |
| Bulk modulus of the ground matrix | $K$ | 174 J/kg |
| Stress coefficient of collagen | $c_3$ | 180 J/kg |
| Uncrimping rate of collagen | $c_4$ | 11 |
| Initial mass density of the ground matrix | $\rho_0^m$ | 500 kg/m³ |
| Initial mass density of meridional collagen | $\rho_0^{fm}$ | 450 kg/m³ |
| Initial mass density of circumferential collagen | $\rho_0^{fc}$ | 50 kg/m³ |
| Growth parameter of collagen | $k_\sigma$ | $\in \{2 \cdot 10^{-4}, 2 \cdot 10^{-3}\}$ days$^{-1}$ |
| Turnover time of collagen | $T$ | 100 days |
| Homeostatic stretch of collagen | $\lambda_h = \lambda_h^{fm} = \lambda_h^{fc}$ | 1.01 |
| **Peripapillary sclera** | | |
| Shear modulus of the ground matrix | $c_1$ | 10 J/kg |
| Bulk modulus of the ground matrix | $K$ | 348 J/kg |
| Stress coefficient of collagen | $c_3$ | 360 J/kg |
| Uncrimping rate of collagen | $c_4$ | 11 |
| Initial mass density of the ground matrix | $\rho_0^m$ | 500 kg/m³ |
| Initial mass density of meridional collagen | $\rho_0^{fm}$ | 50 kg/m³ |
| Initial mass density of circumferential collagen | $\rho_0^{fc}$ | 450 kg/m³ |
| Growth parameter of collagen | $k_\sigma$ | $\in \{2 \cdot 10^{-4}, 2 \cdot 10^{-3}\}$ days$^{-1}$ |
| Turnover time of collagen | $T$ | 100 days |
| Homeostatic stretch of collagen | $\lambda_h = \lambda_h^{fm} = \lambda_h^{fc}$ | 1.01 |
| **Peripheral sclera** | | |
| Shear modulus of the ground matrix | $c_1$ | 10 J/kg |
| Bulk modulus of the ground matrix | $K$ | 348 J/kg |
| Stress coefficient of collagen | $c_3$ | 360 J/kg |
| Uncrimping rate of collagen | $c_4$ | 11 |
| Initial mass density of the ground matrix | $\rho_0^m$ | 500 kg/m³ |
| Initial mass density of meridional collagen | $\rho_0^{fm}$ | 250 kg/m³ |
| Initial mass density of circumferential collagen | $\rho_0^{fc}$ | 250 kg/m³ |
| Growth parameter of collagen | $k_\sigma$ | $\in \{2 \cdot 10^{-4}, 2 \cdot 10^{-3}\}$ days$^{-1}$ |
| Turnover time of collagen | T | 100 days |
| Homeostatic stretch of collagen | $\lambda_h = \lambda_h^{fm} = \lambda_h^{fc}$ | 1.01 |

# Appendix

## Total Cauchy Stress

The total Cauchy stress of the constrained mixture results in

$$\boldsymbol{\sigma} = \frac{1}{J}\mathbf{F}\mathbf{S}\mathbf{F}^{\mathrm{T}} = \frac{2}{J}\mathbf{F}\frac{\partial \Psi}{\partial \mathbf{C}}\mathbf{F}^{\mathrm{T}} \tag{1}$$

$$\boldsymbol{\sigma} = \frac{2}{J}\rho_0^m(t)\mathbf{F}\frac{\partial \overline{W}^m}{\partial \mathbf{C}}\mathbf{F}^{\mathrm{T}} + \frac{2}{J}\rho_0^m(t)\mathbf{F}\frac{\partial K(J_e^m - 1)^2}{\partial \mathbf{C}}\mathbf{F}^{\mathrm{T}} + \frac{2}{J}\rho_0^f(t)\mathbf{F}\frac{\partial W^f}{\partial \mathbf{C}}\mathbf{F}^{\mathrm{T}} \tag{2}$$

$$\begin{aligned}\boldsymbol{\sigma} = \frac{2}{J}\rho_0^m(t)\mathbf{F}\frac{\partial \overline{W}^m}{\partial \mathbf{C}_e}:\frac{\partial \mathbf{C}_e}{\partial \mathbf{C}}\mathbf{F}^{\mathrm{T}} + \frac{2}{J}\rho_0^m(t)\mathbf{F}\frac{\partial K(J_e^m - 1)^2}{\partial \mathbf{C}_e}:\frac{\partial \mathbf{C}_e}{\partial \mathbf{C}}\mathbf{F}^{\mathrm{T}} \\ + \frac{2}{J}\rho_0^f(t)\mathbf{F}\frac{\partial W^f}{\partial \mathbf{C}_e}:\frac{\partial \mathbf{C}_e}{\partial \mathbf{C}}\mathbf{F}^{\mathrm{T}}\end{aligned} \tag{3}$$

The 4$^{\text{th}}$ order tensor $\frac{\partial \mathbf{C}_e}{\partial \mathbf{C}}$ can be expressed as

$$\frac{\partial \mathbf{C}_e}{\partial \mathbf{C}} = \frac{\partial \left(\mathbf{F}_{gr}^{-\mathrm{T}}\mathbf{C}\mathbf{F}_{gr}^{-1}\right)_{ij}}{\partial \mathbf{C}_{kl}} = \frac{1}{2}\left(\mathbf{F}_{gr}^{-\mathrm{T}}\underline{\otimes}\mathbf{F}_{gr}^{-\mathrm{T}} + \mathbf{F}_{gr}^{-\mathrm{T}}\overline{\otimes}\mathbf{F}_{gr}^{-\mathrm{T}}\right) \tag{4}$$

with

$$\left(\mathbf{A}\underline{\otimes}\mathbf{B}\right)_{ijkl} = A_{ik}B_{jl} \tag{5}$$

and

$$\left(\mathbf{A}\overline{\otimes}\mathbf{B}\right)_{ijkl} = A_{il}B_{jk}. \tag{6}$$

Here, we assumed that $\mathbf{F}_{gr} = \mathbf{F}_g\mathbf{F}_r$ is symmetric, a condition that holds true based on the assumptions we have made. Using equation ( 4 ) and assuming that $\mathbf{A}$ is a 2$^{\text{nd}}$ order symmetric tensor, we can show that

$$\mathbf{F}\left(\mathbf{A}:\frac{\partial \mathbf{C}_e}{\partial \mathbf{C}}\right)\mathbf{F}^{\mathrm{T}} = \mathbf{F}_e\mathbf{A}\mathbf{F}_e^{\mathrm{T}} \tag{7}$$

which then results in

$$\boldsymbol{\sigma} = \frac{2}{J}\rho_0^m(t)\mathbf{F}_e\frac{\partial \overline{W}^m}{\partial \mathbf{C}_e}\mathbf{F}_e^{\mathrm{T}} + \frac{2}{J}\rho_0^m(t)\mathbf{F}_e\frac{\partial K(J_e^m - 1)^2}{\partial \mathbf{C}_e}\mathbf{F}_e^{\mathrm{T}} + \frac{2}{J}\rho_0^f(t)\mathbf{F}_e\frac{\partial W^f}{\partial \mathbf{C}_e}\mathbf{F}_e^{\mathrm{T}} \tag{8}$$

**Ground Matrix – Isochoric Contribution to the Cauchy Stress**

The Cauchy stress resulting from the isochoric part of the strain energy function of the ground matrix $\overline{W}^m$ is given as

$$\overline{\boldsymbol{\sigma}}^m = \frac{2}{J}\rho_0^m(t)\mathbf{F}_e \frac{\partial \overline{W}^m}{\partial \mathbf{C}_e}\mathbf{F}_e^T \quad (9)$$

using

$$\frac{\partial \overline{W}^m}{\partial \mathbf{C}_e} = \frac{\partial \overline{W}^m}{\partial \overline{\mathbf{C}}_e} : \frac{\partial \overline{\mathbf{C}}_e}{\partial \mathbf{C}_e}$$

$$= J_e^{-2/3}\left[\frac{\partial \overline{W}^m}{\partial \overline{\mathbf{C}}_e} - \frac{1}{3}\left(\frac{\partial \overline{W}^m}{\partial \overline{\mathbf{C}}_e} : \mathbf{C}_e\right)\mathbf{C}_e^{-1}\right] \quad (10)$$

with

$$\frac{\partial \mathrm{tr}(\overline{\mathbf{C}}_e)}{\partial \overline{\mathbf{C}}_e} = \mathbf{I} \quad (11)$$

we can evaluate the following derivate

$$\frac{\partial \overline{W}^m}{\partial \overline{\mathbf{C}}_e} = \frac{\partial \overline{W}^m}{\partial \overline{I}_{1e}} \frac{\partial \overline{I}_{1e}}{\partial \overline{\mathbf{C}}_e} = c_1 \mathbf{I} \quad (12)$$

so that

$$\frac{\partial \overline{W}^m}{\partial \mathbf{C}_e} = J_e^{-2/3}\left(c_1\mathbf{I} - \frac{1}{3}(c_1\mathbf{I}:\mathbf{C}_e)\mathbf{C}_e^{-1}\right) \quad (13)$$

with the 2$^{\text{nd}}$ order identity tensor $\mathbf{I}$ and

$$\mathbf{I}:\mathbf{C}_e = I_{1e} \quad (14)$$

gives us

$$\frac{\partial \overline{W}^m}{\partial \mathbf{C}_e} = J_e^{-2/3}c_1\left(\mathbf{I} - \frac{I_{1e}}{3}\mathbf{C}_e^{-1}\right). \quad (15)$$

Then, the isochoric contribution of the ground matrix Cauchy stress results in

$$\overline{\boldsymbol{\sigma}}^m = \frac{2}{J}\rho_0^m(t)\mathbf{F}_e \frac{\partial \overline{W}^m}{\partial \mathbf{C}_e}\mathbf{F}_e^T \quad (16)$$

$$= \frac{2}{J}\rho_0^m(t)J_e^{-2/3}c_1\left(J_e^{2/3}\bar{\mathbf{F}}_e\bar{\mathbf{F}}_e^T - \frac{\bar{I}_{1e}}{3}J_e^{2/3}\mathbf{F}_e\mathbf{C}_e^{-1}\mathbf{F}_e^T\right)$$

$$= \frac{2}{J}\rho_0^m(t)c_1\left(\bar{\mathbf{B}}_e - \frac{\bar{I}_{1e}}{3}\mathbf{I}\right)$$

with the isochoric elastic left Cauchy-Green deformation tensor $\bar{\mathbf{B}}_e = \bar{\mathbf{F}}_e\bar{\mathbf{F}}_e^T$.

**Ground Matrix – Volumetric Contribution to the Cauchy Stress**

The Cauchy stress resulting from the volumetric part of the strain energy function of the ground matrix is given as

$$\boldsymbol{\sigma}_{vol}^m = \frac{2}{J}\rho_0^m(t)K\mathbf{F}_e\frac{\partial[(J_e-1)^2]}{\partial \mathbf{C}_e}\mathbf{F}_e^T$$

$$= \frac{2}{J}\rho_0^m(t)K\mathbf{F}_e\frac{\partial[(J_e-1)^2]}{\partial J_e}\frac{\partial J_e}{\partial \mathbf{C}_e}\mathbf{F}_e^T \qquad (17)$$

$$= \frac{2}{J}\rho_0^m(t)K\mathbf{F}_e 2(J_e-1)\frac{J_e}{2}\mathbf{C}_e^{-1}\mathbf{F}_e^T$$

$$= \frac{2}{J}\rho_0^m(t)K(J_e-1)J_e\mathbf{I}.$$

Note, that the hydrostatic pressure p can be defined as

$$p = \frac{\partial K(J_e-1)^2}{\partial J_e} = 2K(J_e-1) \qquad (18)$$

so that the volumetric Cauchy stress becomes

$$\boldsymbol{\sigma}_{vol}^m = \frac{J_e}{J}\rho_0^m(t)p\mathbf{I} \qquad (19)$$

**Collagen – Cauchy Stress**

The Cauchy stress of the collagen fibers results in

$$\boldsymbol{\sigma}^f = \frac{2}{J}\rho_0^f(t)\mathbf{F}_e\frac{\partial W^f(\mathbf{C}_e)}{\partial \mathbf{C}_e}\mathbf{F}_e^T \qquad (20)$$

$$= \frac{2}{J} \rho_0^f(t) \mathbf{F}_e \left( \frac{\partial W^f}{\partial I_{4e}^f} \frac{\partial I_{4e}^f}{\partial \mathbf{C}_e} \right) \mathbf{F}_e^T$$

$$= \frac{2}{J} \rho_0^f(t) \mathbf{F}_e \left( \frac{\partial W^f}{\partial I_{4e}^f} \mathbf{a}_{gr}^f \otimes \mathbf{a}_{gr}^f \right) \mathbf{F}_e^T$$

with

$$\mathbf{F}_e (\mathbf{a}_{gr}^f \otimes \mathbf{a}_{gr}^f) \mathbf{F}_e^T = I_{4e}^f \mathbf{a}^f \otimes \mathbf{a}^f. \tag{21}$$

Here, $\mathbf{a}^f$ represents the fiber direction of the respective collagen fiber family in the current configuration. Furthermore, the scalar derivative $\frac{\partial W^f}{\partial I_{4e}^f}$ evaluates to

$$\frac{\partial W^f}{\partial I_{4e}^f} = c_3 (I_{4e}^f - 1) e^{c_4 (I_{4e}^f - 1)^2}. \tag{22}$$

Then, the Cauchy stress of collagen results in

$$\boldsymbol{\sigma}^f = \frac{2}{J} \rho_0^f(t) c_3 I_{4e}^f (I_{4e}^f - 1) e^{c_4 (I_{4e}^f - 1)^2} \mathbf{a}^f \otimes \mathbf{a}^f \tag{23}$$

**Total Spatial Elasticity Tensor**

The spatial elasticity tensor is given as

$$\mathbb{c} = \frac{1}{J} (\mathbf{F} \underline{\otimes} \mathbf{F}) : \mathbb{C} : (\mathbf{F}^T \underline{\otimes} \mathbf{F}^T) \tag{24}$$

with the material elasticity tensor

$$\mathbb{C} = 2 \frac{\partial \mathbf{S}}{\partial \mathbf{C}}. \tag{25}$$

Here, $\mathbf{S}$ is the second Piola-Kirchhoff stress tensor.

**Ground Matrix – Isochoric Contribution to the Spatial Elasticity Tensor**

The second Piola-Kirchhoff stress resulting from the isochoric part of the strain energy function of the ground matrix $\overline{W}^m$ is given as

$$\overline{\mathbf{S}}^m = J \mathbf{F}^{-1} \overline{\boldsymbol{\sigma}}^m \mathbf{F}^{-T} \tag{26}$$

$$= 2\rho_0^m(t)c_1 \mathbf{F}^{-1}\left(\bar{\mathbf{B}}_e - \frac{\bar{I}_{1e}}{3}\mathbf{I}\right)\mathbf{F}^{-T}$$

$$= 2\rho_0^m(t)c_1 \left(\mathbf{F}^{-1}\bar{\mathbf{B}}_e \mathbf{F}^{-T} - \frac{\bar{I}_{1e}}{3}\mathbf{C}^{-1}\right)$$

$$= 2\rho_0^m(t)c_1 J_e^{-2/3}\left(\mathbf{C}_{gr}^{-1} - \frac{I_{1e}}{3}\mathbf{C}^{-1}\right)$$

with the inverse right Cauchy-Green deformation tensor $\mathbf{C}_{gr}^{-1} = \mathbf{F}_{gr}^{-1}\mathbf{F}_{gr}^{-T}$. Then, the material elasticity tensor results in

$$\bar{\mathbb{C}}^m = 2\frac{\partial \bar{\mathbf{S}}^m}{\partial \mathbf{C}}$$

$$= 4\rho_0^m(t)c_1 \left[\frac{\partial(J_e^{-2/3}\mathbf{C}_{gr}^{-1})}{\partial \mathbf{C}} - \frac{1}{3}\frac{\partial(J_e^{-2/3}I_{1e}\mathbf{C}^{-1})}{\partial \mathbf{C}}\right] \quad (27)$$

The first term in the equation (27) results in

$$\frac{\partial(J_e^{-2/3}\mathbf{C}_{gr}^{-1})}{\partial \mathbf{C}} = \mathbf{C}_{gr}^{-1} \otimes \frac{\partial J_e^{-2/3}}{\partial \mathbf{C}} + J_e^{-2/3}\frac{\partial \mathbf{C}_{gr}^{-1}}{\partial \mathbf{C}}$$

$$= \mathbf{C}_{gr}^{-1} \otimes \frac{\partial J_e^{-2/3}}{\partial J_e}\frac{\partial J_e}{\partial \mathbf{C}_e}:\frac{\partial \mathbf{C}_e}{\partial \mathbf{C}} + J_e^{-2/3}\frac{\partial \mathbf{C}_{gr}^{-1}}{\partial \mathbf{C}} \quad (28)$$

$$= -\frac{2}{3}J_e^{-5/3}\mathbf{C}_{gr}^{-1} \otimes \frac{J_e}{2}\mathbf{C}^{-1} + J_e^{-2/3}\frac{\partial \mathbf{C}_{gr}^{-1}}{\partial \mathbf{C}}$$

$$= -\frac{1}{3}J_e^{-2/3}\mathbf{C}_{gr}^{-1} \otimes \mathbf{C}^{-1} + J_e^{-2/3}\frac{\partial \mathbf{C}_{gr}^{-1}}{\partial \mathbf{C}}$$

Assuming an explicit time integration scheme to solve the evolution equations for $\mathbf{F}_{gr}$ ($\mathbf{F}_{gr}$ is constant during each time increment), the second summand in equation (28) results in

$$J_e^{-2/3}\frac{\partial \mathbf{C}_{gr}^{-1}}{\partial \mathbf{C}} = \mathbb{0}. \quad (29)$$

The second term in equation (27) evaluates to

$$\frac{\partial(J_e^{-2/3}I_{1e}\mathbf{C}^{-1})}{\partial \mathbf{C}} = \mathbf{C}^{-1} \otimes \frac{\partial J_e^{-2/3}I_{1e}}{\partial \mathbf{C}} + J_e^{-2/3}I_{1e}\frac{\partial \mathbf{C}^{-1}}{\partial \mathbf{C}} \quad (30)$$

$$= \mathbf{C}^{-1} \otimes \frac{\partial J_e^{-2/3} I_{1e}}{\partial \mathbf{C}} - J_e^{-2/3} I_{1e} \mathbf{C}^{-1} \odot \mathbf{C}^{-1}$$

$$= \mathbf{C}^{-1} \otimes \frac{\partial J_e^{-2/3} I_{1e}}{\partial \mathbf{C}_e} : \frac{\partial \mathbf{C}_e}{\partial \mathbf{C}} - J_e^{-2/3} I_{1e} \mathbf{C}^{-1} \odot \mathbf{C}^{-1}$$

$$= \mathbf{C}^{-1} \otimes \left( J_e^{-2/3} \frac{\partial I_{1e}}{\partial \mathbf{C}_e} + I_{1e} \frac{\partial J_e^{-2/3}}{\partial \mathbf{C}_e} \right) : \frac{\partial \mathbf{C}_e}{\partial \mathbf{C}} - J_e^{-2/3} I_{1e} \mathbf{C}^{-1} \odot \mathbf{C}^{-1}$$

$$= \mathbf{C}^{-1} \otimes \left( J_e^{-2/3} \frac{\partial I_{1e}}{\partial \mathbf{C}_e} - \frac{1}{3} I_{1e} J_e^{-2/3} \mathbf{C}_e^{-1} \right) : \frac{\partial \mathbf{C}_e}{\partial \mathbf{C}} - J_e^{-2/3} I_{1e} \mathbf{C}^{-1} \odot \mathbf{C}^{-1}$$

$$= \mathbf{C}^{-1} \otimes \left( J_e^{-2/3} \mathbf{I} - \frac{1}{3} I_{1e} J_e^{-2/3} \mathbf{C}_e^{-1} \right) : \frac{\partial \mathbf{C}_e}{\partial \mathbf{C}} - J_e^{-2/3} I_{1e} \mathbf{C}^{-1} \odot \mathbf{C}^{-1}$$

$$= \mathbf{C}^{-1} \otimes \left( J_e^{-2/3} \mathbf{I} : \frac{\partial \mathbf{C}_e}{\partial \mathbf{C}} \right) - \frac{1}{3} I_{1e} J_e^{-2/3} \mathbf{C}^{-1} \otimes \mathbf{C}^{-1} - J_e^{-2/3} I_{1e} \mathbf{C}^{-1} \odot \mathbf{C}^{-1}$$

$$= J_e^{-2/3} \mathbf{C}^{-1} \otimes \mathbf{C}_{gr}^{-1} - \frac{1}{3} I_{1e} J_e^{-2/3} \mathbf{C}^{-1} \otimes \mathbf{C}^{-1} - J_e^{-2/3} I_{1e} \mathbf{C}^{-1} \odot \mathbf{C}^{-1}$$

with the 4$^{th}$ order tensor

$$(\mathbf{C}^{-1} \odot \mathbf{C}^{-1})_{ijkl} = \frac{1}{2} \left( C_{ik}^{-1} C_{jl}^{-1} + C_{il}^{-1} C_{jk}^{-1} \right). \quad (31)$$

Assembling both terms, the material elasticity tensor results in

$$\begin{aligned}\bar{\mathbb{C}}^m &= 4\rho_0^m(t) c_1 \left[ -\frac{1}{3} J_e^{-2/3} \mathbf{C}_{gr}^{-1} \otimes \mathbf{C}^{-1} \right. \\
&\quad \left. -\frac{1}{3} \left( J_e^{-2/3} \mathbf{C}^{-1} \otimes \mathbf{C}_{gr}^{-1} - \frac{1}{3} I_{1e} J_e^{-2/3} \mathbf{C}^{-1} \otimes \mathbf{C}^{-1} - J_e^{-2/3} I_{1e} \mathbf{C}^{-1} \right. \right. \\
&\quad \left. \left. \odot \mathbf{C}^{-1} \right) \right] \\
&= \frac{4}{3} \rho_0^m(t) c_1 J_e^{-2/3} \left( \frac{1}{3} I_{1e} \mathbf{C}^{-1} \otimes \mathbf{C}^{-1} + I_{1e} \mathbf{C}^{-1} \odot \mathbf{C}^{-1} - \mathbf{C}_{gr}^{-1} \otimes \mathbf{C}^{-1} \right. \\
&\quad \left. - \mathbf{C}^{-1} \otimes \mathbf{C}_{gr}^{-1} \right).\end{aligned} \quad (32)$$

We can then calculate the spatial elasticity tensor

$$\begin{aligned}\bar{\mathbb{c}}^m &= \frac{1}{J} (\mathbf{F} \underline{\otimes} \mathbf{F}) : \bar{\mathbb{C}}^m : (\mathbf{F}^T \underline{\otimes} \mathbf{F}^T) \\
&= \frac{4}{3J} \rho_0^m(t) c_1 J_e^{-2/3} \left( \frac{1}{3} I_{1e} \mathbf{I} \otimes \mathbf{I} + I_{1e} \mathbb{S} - \mathbf{B}_e \otimes \mathbf{I} - \mathbf{I} \otimes \mathbf{B}_e \right)\end{aligned} \quad (33)$$

$$= \frac{4}{3J}\rho_0^m(t)c_1\left(\frac{1}{3}\bar{I}_{1e}\mathbf{I}\otimes\mathbf{I} + \bar{I}_{1e}\mathbb{S} - \bar{\mathbf{B}}_e\otimes\mathbf{I} - \mathbf{I}\otimes\bar{\mathbf{B}}_e\right)$$

with the special 4$^{th}$ order identity tensor

$$\mathbb{S}_{ijkl} = \frac{1}{2}(\delta_{ik}\delta_{jl} + \delta_{il}\delta_{jk}). \tag{34}$$

However, in Abaqus, the spatial elasticity tensor is defined in terms of the Jaumann rate of the Cauchy stress, denoted here as $\bar{\mathbb{c}}^m_{jau}$. To obtain the spatial elasticity tensor required by Abaqus, we have to add two terms that depend on the current Cauchy stress of the material

$$\bar{\mathbb{c}}^m_{jau} = \bar{\mathbb{c}}^m + \bar{\boldsymbol{\sigma}}^m \odot \mathbf{I} + \mathbf{I} \odot \bar{\boldsymbol{\sigma}}^m \tag{35}$$

Using equation (16) in (35) results in

$$\bar{\mathbb{c}}^m_{jau} = \frac{4}{3J}\rho_0^m(t)c_1\left(\frac{1}{3}\bar{I}_{1e}\mathbf{I}\otimes\mathbf{I} - \bar{\mathbf{B}}_e\otimes\mathbf{I} - \mathbf{I}\otimes\bar{\mathbf{B}}_e \right. \\ \left. + \frac{3}{2}(\mathbf{I}\underline{\otimes}\bar{\mathbf{B}}_e + \bar{\mathbf{B}}_e\underline{\otimes}\mathbf{I} + \mathbf{I}\overline{\otimes}\bar{\mathbf{B}}_e + \bar{\mathbf{B}}_e\overline{\otimes}\mathbf{I})\right) \tag{36}$$

**Ground Matrix – Volumetric Contribution to the Spatial Elasticity Tensor**

The second Piola-Kirchhoff stress resulting from the volumetric part of the strain energy function of the ground matrix is given as

$$\mathbf{S}^m_{vol} = 2\rho_0^m(t)K\frac{\partial[(J_e-1)^2]}{\partial\mathbf{C}}$$

$$= 2\rho_0^m(t)K\frac{\partial[(J_e-1)^2]}{\partial\mathbf{C}_e}:\frac{\partial\mathbf{C}_e}{\partial\mathbf{C}}$$

$$= 2\rho_0^m(t)K\frac{\partial[(J_e-1)^2]}{\partial J_e}\frac{\partial J_e}{\partial\mathbf{C}_e}:\frac{\partial\mathbf{C}_e}{\partial\mathbf{C}}$$

$$= 2\rho_0^m(t)K(J_e-1)J_e\mathbf{C}_e^{-1}:\frac{\partial\mathbf{C}_e}{\partial\mathbf{C}} \tag{37}$$

$$= 2\rho_0^m(t)K(J_e-1)J_e\mathbf{F}_e^{-1}\mathbf{F}_e^{-T}:\frac{\partial\mathbf{C}_e}{\partial\mathbf{C}}$$

$$= 2\rho_0^m(t)K(J_e-1)J_e\mathbf{F}_{gr}\mathbf{F}^{-1}\mathbf{F}^{-T}\mathbf{F}_{gr}^T:\frac{\partial\mathbf{C}_e}{\partial\mathbf{C}}$$

$$= 2\rho_0^m(t)K(J_e-1)J_e\mathbf{F}_{gr}\mathbf{C}^{-1}\mathbf{F}_{gr}^T:\frac{\partial\mathbf{C}_e}{\partial\mathbf{C}}$$

Using equation ( 4 ), the double dot product in equation (37) results in

$$\mathbf{F}_{gr}\mathbf{C}^{-1}\mathbf{F}_{gr}^T : \frac{\partial \mathbf{C}_e}{\partial \mathbf{C}} = \mathbf{C}^{-1} \tag{38}$$

so that

$$\mathbf{S}_{vol}^m = 2\rho_0^m(t)K(J_e - 1)J_e\mathbf{C}^{-1}$$
$$= \rho_0^m(t)J_e p \mathbf{C}^{-1} . \tag{39}$$

Therefore

$$\mathbb{C}_{vol}^m = 2\frac{\partial \mathbf{S}_{vol}^m}{\partial \mathbf{C}}$$
$$= 2\rho_0^m(t)\frac{\partial (J_e p \mathbf{C}^{-1})}{\partial \mathbf{C}}$$
$$= 2\rho_0^m(t)\mathbf{C}^{-1} \otimes \frac{\partial J_e p}{\partial \mathbf{C}} + 2\rho_0^m(t)J_e p \frac{\partial \mathbf{C}^{-1}}{\partial \mathbf{C}}$$
$$= 2\rho_0^m(t)\mathbf{C}^{-1} \otimes \frac{\partial J_e p}{\partial \mathbf{C}} - 2\rho_0^m(t)J_e p \mathbf{C}^{-1} \odot \mathbf{C}^{-1} \tag{40}$$

with

$$\frac{\partial J_e p}{\partial \mathbf{C}} = \frac{\partial J_e p}{\partial J_e}\frac{\partial J_e}{\partial \mathbf{C}_e}:\frac{\partial \mathbf{C}_e}{\partial \mathbf{C}}$$
$$= K(2J_e - 1)J_e \mathbf{C}_e^{-1}:\frac{\partial \mathbf{C}_e}{\partial \mathbf{C}} \tag{41}$$
$$= K(2J_e - 1)J_e \mathbf{C}^{-1}$$

Then, the material elasticity tensor results in

$$\mathbb{C}_{vol}^m = 2\rho_0^m(t)K(2J_e - 1)J_e\mathbf{C}^{-1} \otimes \mathbf{C}^{-1} - 2\rho_0^m(t)J_e p \mathbf{C}^{-1} \odot \mathbf{C}^{-1}$$
$$= 2\rho_0^m(t)K(2J_e - 1)J_e\mathbf{C}^{-1} \otimes \mathbf{C}^{-1} - 2\rho_0^m(t)J_e 2K(J_e - 1)\mathbf{C}^{-1} \odot \mathbf{C}^{-1} \tag{42}$$
$$= 4\rho_0^m(t)KJ_e\left[\left(J_e - \frac{1}{2}\right)\mathbf{C}^{-1} \otimes \mathbf{C}^{-1} - (J_e - 1)\mathbf{C}^{-1} \odot \mathbf{C}^{-1}\right].$$

Substituting equation (42) into equation (24), the spatial elasticity tensor evaluates to

$$\mathbb{C}_{\text{vol}}^{\text{m}} = 2\rho_0^{\text{m}}(t)K(2J_e - 1)\frac{J_e}{J}\mathbf{I}\otimes\mathbf{I} - 2\rho_0^{\text{m}}(t)\frac{J_e}{J}p\mathbb{S}$$

$$= 2\rho_0^{\text{m}}(t)K\frac{J_e}{J}[(2J_e - 1)\mathbf{I}\otimes\mathbf{I} - 2(J_e - 1)\mathbb{S}] \tag{43}$$

$$= 4\rho_0^{\text{m}}(t)K\frac{J_e}{J}\left[\left(J_e - \frac{1}{2}\right)\mathbf{I}\otimes\mathbf{I} - (J_e - 1)\mathbb{S}\right]$$

Inserting equation (43) into equation (35), the Jaumann rate of the Cauchy stress (i.e. the spatial elasticity tensor required by Abaqus) results in

$$\mathbb{C}_{\text{vol,jau}}^{\text{m}} = 4\rho_0^{\text{m}}(t)K\frac{J_e}{J}\left(J_e - \frac{1}{2}\right)\mathbf{I}\otimes\mathbf{I} \tag{44}$$

**Collagen - Spatial Elasticity Tensor**

The second Piola-Kirchhoff stress of collagen results in

$$\mathbf{S}^{\text{f}} = \frac{2\rho_0^{\text{f}}(t)c_3\left(I_{4e}^{\text{f}} - 1\right)e^{c_4\left(I_{4e}^{\text{f}}-1\right)^2}}{\left\|\mathbf{F}_{\text{gr}}\mathbf{a}_0^{\text{f}}\right\|^2}\mathbf{a}_0^{\text{f}}\otimes\mathbf{a}_0^{\text{f}} \tag{45}$$

with the fiber direction $\mathbf{a}_0^{\text{f}}$ in reference configuration. Then, the material elasticity tensor is given as

$$\mathbb{C}^{\text{f}} = 2\frac{\partial \mathbf{S}^{\text{f}}}{\partial \mathbf{C}}$$

$$= \frac{4\rho_0^{\text{f}}(t)c_3\left[1 + 2c_4\left(I_{4e}^{\text{f}} - 1\right)^2\right]e^{c_4\left(I_{4e}^{\text{f}}-1\right)^2}}{\left\|\mathbf{F}_{\text{gr}}\mathbf{a}_0^{\text{f}}\right\|^4}\mathbf{a}_0^{\text{f}}\otimes\mathbf{a}_0^{\text{f}}\otimes\mathbf{a}_0^{\text{f}}\otimes\mathbf{a}_0^{\text{f}} \tag{46}$$

Substituting equation (46) in equation (24), the spatial elasticity tensor results in

$$\mathbb{c}^{\text{f}} = \frac{4}{J}\rho_0^{\text{f}}(t)c_3\left(I_{4e}^{\text{f}}\right)^2\left[1 + 2c_4\left(I_{4e}^{\text{f}} - 1\right)^2\right]e^{c_4\left(I_{4e}^{\text{f}}-1\right)^2}\mathbf{a}^{\text{f}}\otimes\mathbf{a}^{\text{f}}\otimes\mathbf{a}^{\text{f}}\otimes\mathbf{a}^{\text{f}} \tag{47}$$

The implementation in Abaqus requires the Jaumann rate of the Cauchy stress that is given as

$$\mathbb{c}_{\text{jau}}^{\text{f}} = \mathbb{c}^{\text{f}} + \boldsymbol{\sigma}^{\text{f}}\odot\mathbf{I} + \mathbf{I}\odot\boldsymbol{\sigma}^{\text{f}}. \tag{48}$$

Equation (48) could not be further simplified and it was directly implemented in Abaqus.